\documentclass[manuscript]{aastex631}

\newcommand{\teff}{T$_{eff}$}

\shorttitle{JWST Spectral Standards}
\shortauthors{Pritchard et al.}

\begin{document}

\title{JWST/NIRSpec Spectral Standards for M, L, \& T Dwarfs and Subdwarfs}

\author[0009-0000-0608-7870]{Evan Pritchard}
\affiliation{UC San Diego, La Jolla, CA 92093, USA}

\author[0009-0001-6464-1160]{Marylin Loritsch}
\affiliation{UC San Diego, La Jolla, CA 92093, USA}

\author[0009-0001-7549-2535]{Julia Haynes}
\affiliation{UC San Diego, La Jolla, CA 92093, USA}

\author[0009-0006-0192-9630]{Sara Morrissey}
\affiliation{University of Notre Dame, IN 46556, USA}

\author[0000-0002-1420-1837]{Emma Softich}
\affiliation{UC San Diego, La Jolla, CA 92093, USA}

\author[0000-0002-6523-9536]{Adam J.~Burgasser}
\affiliation{UC San Diego, La Jolla, CA 92093, USA}

\begin{abstract}
We present low-resolution infrared spectral standards in the 0.76--5.0 $\mu$m range based on JWST NIRspec Prism data from deep sky surveys. Our standards encompass spectral types M, L, and T, and dwarf and subdwarf metallicity classes.
The standards extend ground-based spectral templates and enable classification of low-temperature stellar and substellar discoveries from deep JWST/NIRSpec data.
\end{abstract}


\section{Introduction} \label{sec:intro}

Ultracool dwarfs (UCDs) encompass the lowest mass stars (M $\lesssim0.1\,M_\odot$) and brown dwarfs (M $\lesssim0.075\,M_\odot$), with effective temperatures {\teff} $\lesssim$ 3500~K. 
An abundant population
%Among the most populous stars 
in the Milky Way ($\sim$~20\% of the 20 pc sample; \citealt{2024ApJS..271...55K}),
UCDs undergo minimal or no hydrogen fusion, giving them exceptionally long lifetimes ($\tau \gtrsim 10^{11}$~yr) and making them potential probes of the early assembly and chemical evolution of the Milky Way.
%\citep{2017arXiv170200091C}.  % problematic reference so skipping
However, these sources are cold and dim, requiring high infrared sensitivity to probe distant and ancient halo and thick disk populations.

%Accordingly, 
%The James Webb Space Telescope (JWST) 
JWST has significantly improved our ability to search for these ancient UCDs, with studies reporting the spectroscopic identification of metal-poor sources out to kiloparsec scales (e.g., \citealt{2024ApJ...962..177B,2024ApJ...975...31H,2026AJ....171..191M}).  
Accurate characterization of these distant sources requires
precise spectral classification; yet, spectral standards for UCDs are currently defined at optical and near-infrared wavelengths (NIR; 0.75--2.5$\mu$m) accessible from the ground. 
%\citep{2006ApJ...637.1067B,2010ApJS..190..100K}.
Here, we present spectral standards covering 0.76--5.0~$\mu$m drawn from JWST/NIRSpec spectra that aim to improve spectral and metallicity classification of distant UCDs.

\section{UCD Spectral Sample} \label{sec:sample}

We identified UCD spectra from the DAWN JWST archive 
\citep{brammer_2023_8370018,2024Sci...384..890H,2025A&A...697A.189D}, 
a public repository of survey data which includes more than 80,000 spectra acquired with JWST/NIRSpec in its Prism-dispersed mode
(0.6 $\leq \lambda \leq$ 5.3~$\mu$m, $\lambda/\Delta\lambda$ = 30--330; \citealt{2022A&A...661A..80J}). 
We selected spectra with low inferred redshifts (z $<$ 0.1) and high spectrum quality (grade $>$ 2), and visually confirmed candidates based on UCD features such as peak emission over 0.9--1.3~$\mu$m and 
absorption bands from H$_2$O and CH$_4$, among others. 
We rejected low signal-to-noise spectra (median S/N $\leq$ 2.3) and sources that appeared to be warmer stars.
In total, we identified 67 UCD objects from DAWN observed as part of the 
BoRG \citep{2025ApJ...983...18R}, 
Nelson \citep{2023ApJ...948L..18N}, 
CEERS \citep{2023ApJ...946L..13F}, 
GTO \citep{2024A&A...689A..73M}, 
JADES \citep{2025ApJS..277....4D}, 
NEXUS \citep{2026ApJS..282...54Z},
RUBIES \citep{2025A&A...697A.189D}, and 
UNCOVER \citep{2024ApJ...974...92B} surveys.

\section{Selection of Standards} \label{sec:spectra}

UCD spectra were ordered by increasing molecular absorption and redward shift of the emission peak, creating a sequence based primarily on temperature. The spectra were then compared to NIR spectral standards for M, L, and T dwarfs 
\citep{2006ApJ...637.1067B,2010ApJS..190..100K} 
and subdwarfs 
\citep{2019AJ....158..182G,2025ApJ...982...79B}.
We identified the best match using the root mean square difference after optimal relative scaling, and assigned that standard's type as the source's spectral classification. We visually confirmed these matches, and spectra showing excellent agreement with NIR standards were selected as JWST standard candidates.
%, and spectra showing clear deviations identified as outliers. 
When more than one JWST spectrum matched a NIR standard, we selected the spectrum with the lowest root mean square residual and/or highest S/N.

\begin{figure}
    \centering
    \includegraphics[width=0.49\linewidth]{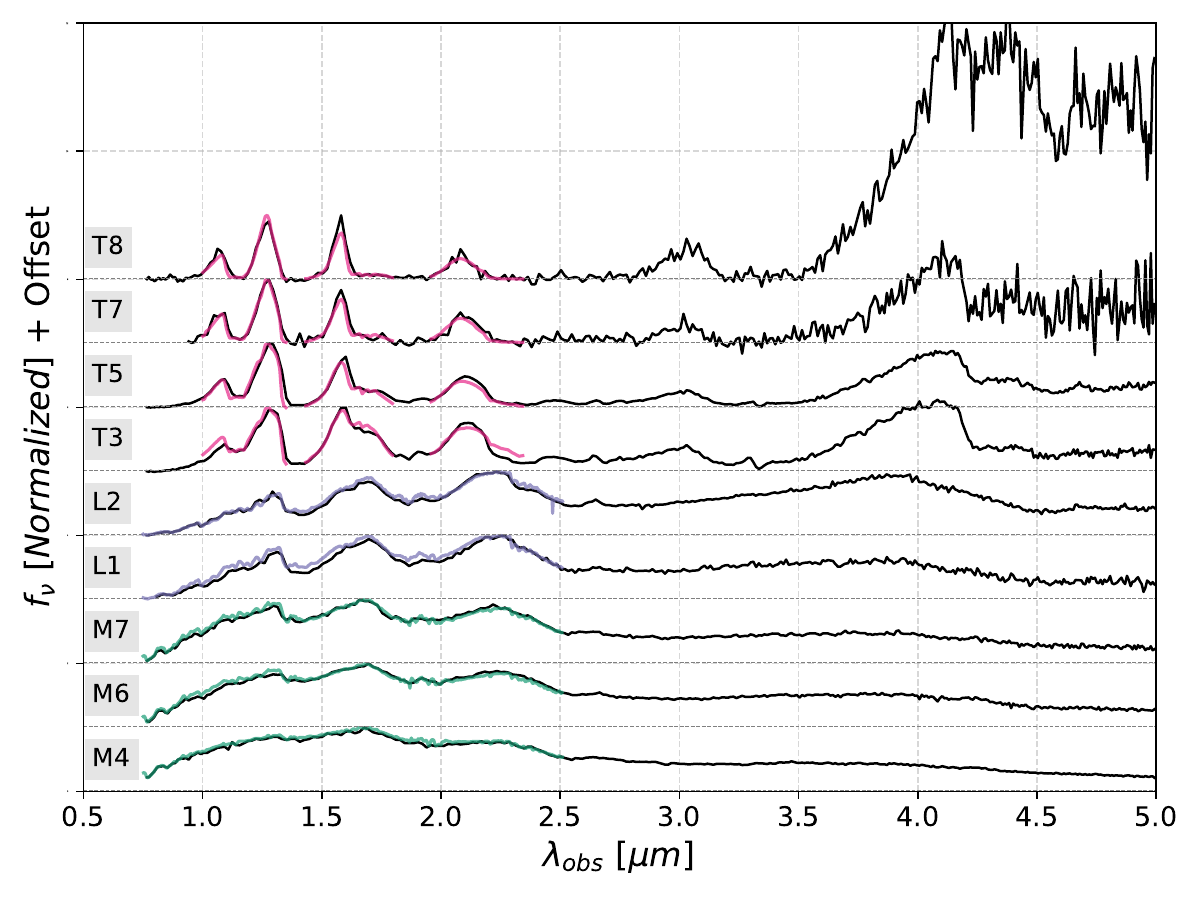}
    \includegraphics[width=0.49\linewidth]{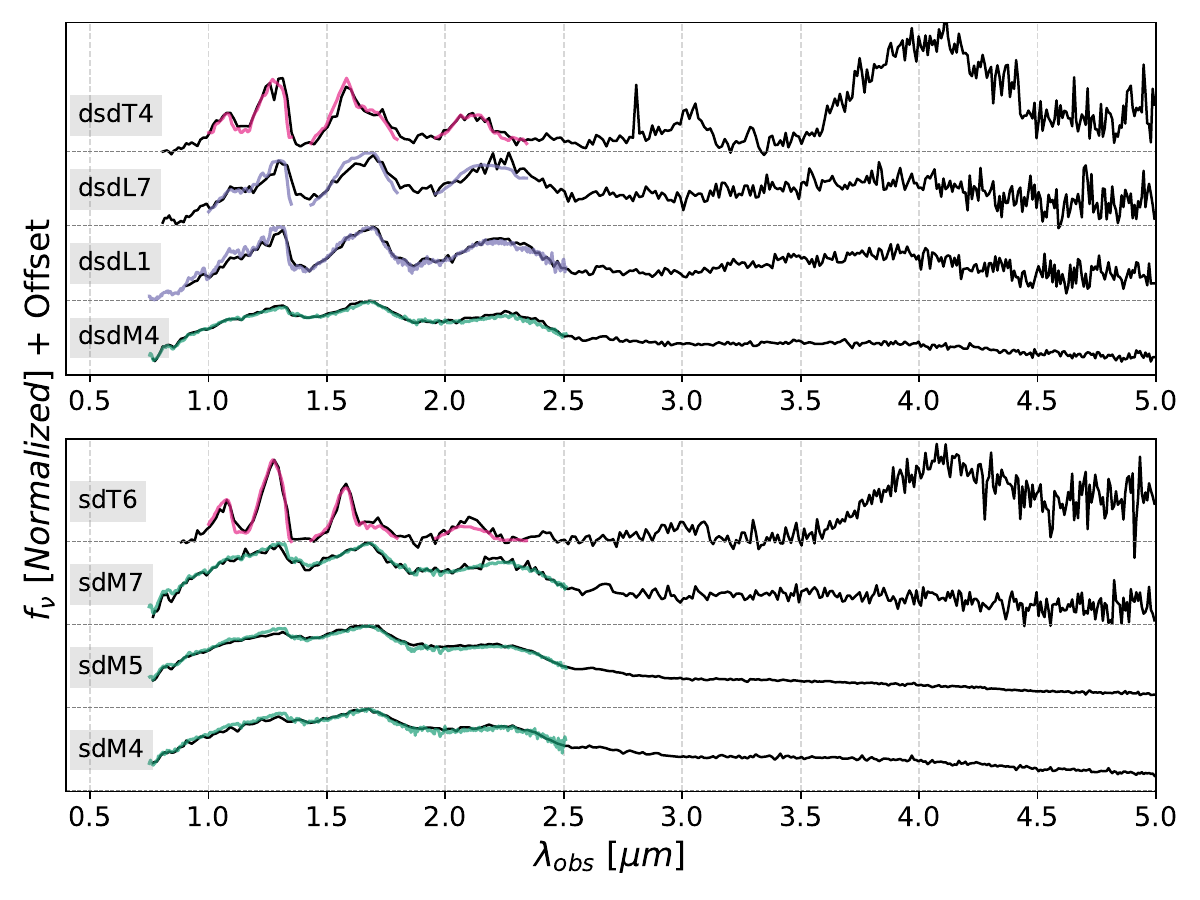} \\
    (a)\hspace{3.2in}(b)\vspace{0.2in} \\
    \includegraphics[width=0.49\linewidth]{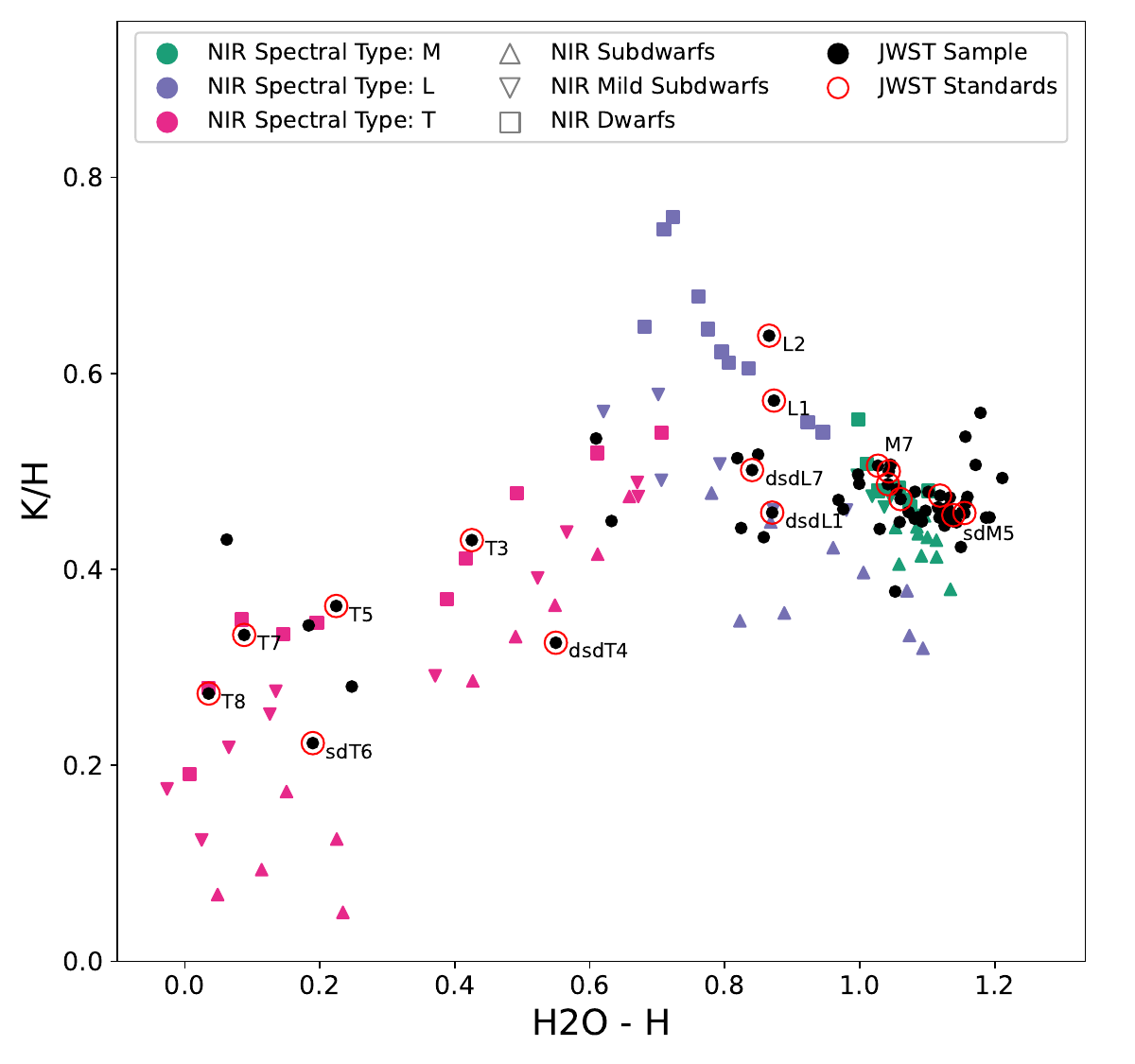}
    \includegraphics[width=0.49\linewidth]{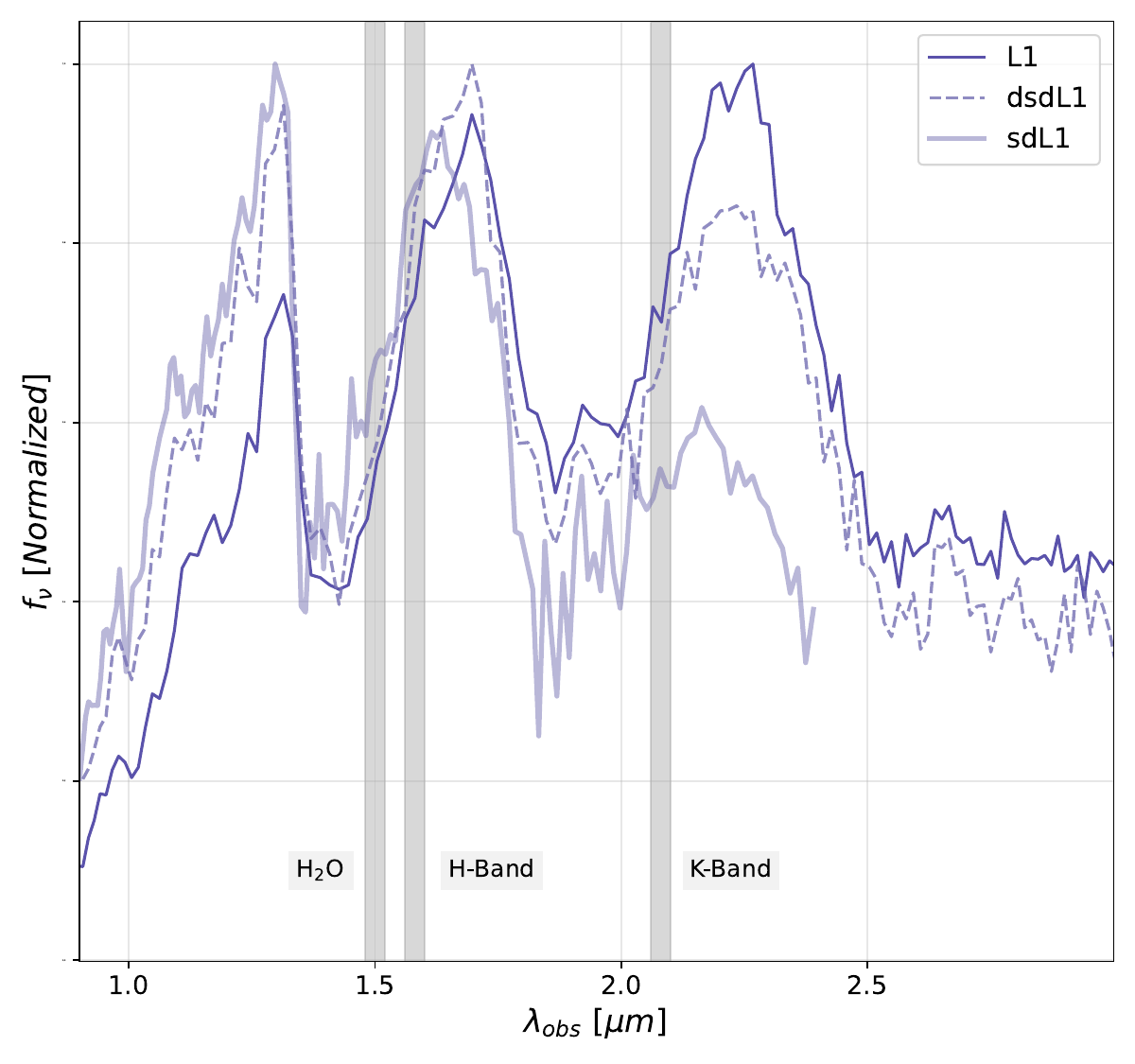} \\
    (c)\hspace{3.2in}(d)
    \caption{
    (a and b) JWST/NIRSpec Prism spectral standards for M, L, and T dwarfs (a) and subdwarfs (b), plotted in $f_\nu$ flux density units (black lines). Spectra are normalized in the 1--2~$\mu$m region and offset by constants. Spectral types based on the best-match NIR standards (overplotted color lines) are listed along the left side.
    %, and corresponding best-fit NIR standards are overplotted (green, blue, and magenta lines).
%    (Top Right) Sequence for MLT subdwarfs and mild subdwarfs-- same format as above.
    (c) Comparison of spectral indices H$_2$O-H and K/H for ground-based spectral standards (colored symbols) and JWST/NIRSpec Prism spectra (black dots).
    Selected NIRSpec standards are encircled and labeled. 
    For the NIR standards, symbol colors indicate spectral class (M dwarf = green, L dwarf = blue, T dwarf = magenta), while symbol shapes indicate metallicity class (dwarf = squares, d/sd = inverted triangle, sd = triangles).
    (d) JWST/NIRSpec Prism spectra of our L1 (solid blue line) and d/sdL1 (dashed blue line) standards, compared to the NIR sdL1 standard (faded blue line; \citealt{2025ApJ...982...79B}) over 0.9--3.0~$\mu$m. Shaded bands indicate the regions sampled for the H$_2$O-H (H$_2$O over H-band) and K/H (K-band over H-band) indices. 
    % H indicating the H$_2$O band (1.48--1.52$\mu$m), the H-band continuum (1.56-1.60$\mu$m), and the K-band continuum (2.06-2.10$\mu$m) which define the indices plotted in (c). 
}
    \label{fig:figure}
\end{figure}

Our selected standards are shown in Figures~\ref{fig:figure}a and b.
They span the full spectral class range from mid-M to late-T, albeit with gaps due to the limited selection of deep-field sources available in DAWN.
There is a large gap between dwarf types L2 and T3, possibly the result of brown dwarf cooling among the older thick disk and halo populations; and only two T subdwarfs (one mild and one normal subdwarf) were identified.
We distinguish between dwarfs, mild subdwarfs (d/sd) and subdwarfs (sd), as no sources matching lower metallicity classes (esd and usd) were found in our sample.

Figure~\ref{fig:figure}c compares measurements of
%the NIR and JWST/NIRSpec standards, and the remainder of our sample, along 
two spectral indices, H$_2$O-H and K/H, which measure the strength of H$_2$O absorption at 1.4~$\mu$m and the relative flux peaks at 2.1 and 1.6~$\mu$m, respectively 
%features at specific wavelength bands within each spectra 
\citep{2006ApJ...637.1067B}. 
These indices were identified for their ability to separate spectral (H$_2$O-H) and metallicity (K/H) classes, particularly for the L and T (sub)dwarfs.
Figure~\ref{fig:figure}d illustrates these differences in the spectra of our L1 and d/sdL1 standards,
%as well as the former NIR sdL1 standard. These standards 
which show similar H$_2$O depths but clear differences in peak flux densities.
%, motivating our differentiation between d/sd and sd standards. 
Other regions over the 0.76--5.0~$\mu$m range, such as the depth of 2.7~$\mu$m band or ratio of 4.0~$\mu$m and 2.1~$\mu$m peaks, may also prove to be useful temperature and metallicity indicators outside the traditional NIR region.

% THE TEXT BELOW HAS BEEN CONDENSED INTO ABOVE
% along the x and y axes, respectively. 
% In Figure~\ref{fig:figure} by plotting H$_2$O-H along the x-axis, T dwarf/subdwarfs exclusively occupy regions with H$_2$O-H indices $\lesssim0.7$, L dwarf/subdwarfs occupy $\geq 0.7$ but $\leq1.0$, while M dwarf/subdwarfs clump around $\sim~1.0$. 
% Within their respective spectral type regimes, T-dwarf/subdwarfs and L-dwarf/subdwarfs additionally separate by metallicity along the K/H index in the y-axis, with dwarfs having consistently higher K/H indices. Recognizing these trends set by spectral indices validates our selection of standards by confirming each selected standard properly resides in their expected spectral and metallicity regime. 

% Figure~\ref{fig:figure} also includes the spectra of JWST/NIRSpec L1 and d/sdL1 standards within the wavelength range 0.9-3.0~$\mu$m. The shaded regions indicate the specific wavelength bands sampled by the spectral indices to help visualize how these indices can distinguish between spectral and metallicity types. Furthermore, the difference between the 1.3~$\mu$m and 2.25~$\mu$m peaks, indicate a significant dependence of spectral features on metallicity, motivating the distinction between mild subdwarfs and subdwarfs in our sample. 

The standard set presented here enables use of
%its kind-- using JWST PRISM spectra to fully utilize 
the full wavelength range available for JWST NIRSpec/Prism spectra for
%0.76-5.0~$\mu$m. This extended wavelength range provides 
improved classification of distant UCDs.
%, especially late-L and T-dwarfs, as their spectra become more pronounced at longer wavelengths. 
%Although lacking a continuous sequence of spectral and metallicity types, 
The present gaps in the sequence are expected to be filled in as deep-field JWST spectroscopic surveys continue discovering UCDs, 
%while larger samples will 
enabling more refined classification along temperature, metallicity, and other characteristics (e.g., surface gravity; \citealt{2013ApJ...772...79A}).
%Nevertheless, these standards provide an opportunity to classify new JWST UCD discoveries 
%along temperature and metallicity axes, necessary for distinguishing Milky Way populations.
%with the resolution and range provided by JWST.
We note that recent results from SPHEREx provide an alternative set of standards over the NIRSpec/Prism range at lower resolution ($\lambda/\Delta\lambda \approx 50$; \citealt{gagne2026}).
The standard spectra are included as digital files with this article.

\begin{acknowledgments}
This research was supported by the STARTastro program, funded by the Heising Simons Foundation and National Science Foundation. Data products presented here were retrieved from the Dawn JWST Archive, an initiative of the Cosmic Dawn Center (DAWN) which is funded by the Danish National Research Foundation under grant DNRF140.
\end{acknowledgments}

%\clearpage
\bibliography{new.ms}{}

@software{brammer_2023_8370018,
  author       = {Brammer, Gabriel},
  title        = {grizli},
  month        = sep,
  year         = 2023,
  publisher    = {Zenodo},
  version      = {1.9.11},
  doi          = {10.5281/zenodo.8370018},
  url          = {https://doi.org/10.5281/zenodo.8370018},
}

@ARTICLE{2006ApJ...637.1067B,
       author = {{Burgasser}, Adam J. and {Geballe}, T.~R. and {Leggett}, S.~K. and {Kirkpatrick}, J. Davy and {Golimowski}, David A. and {others}},
        title = "{A Unified Near-Infrared Spectral Classification Scheme for T Dwarfs}",
      journal = {\apj},
         year = 2006,
        month = feb,
       volume = {637},
       number = {2},
        pages = {1067-1093}
}

@ARTICLE{2024ApJ...962..177B,
       author = {{Burgasser}, Adam J. and {Bezanson}, Rachel and {Labbe}, Ivo and {Brammer}, Gabriel and {Cutler}, Sam E. and {Furtak}, Lukas J. and {Greene}, Jenny E. and {Gerasimov}, Roman and {Leja}, Joel and {Pan}, Richard and {Price}, Sedona H. and {Wang}, Bingjie and {Weaver}, John R. and {Whitaker}, Katherine E. and {Fujimoto}, Seiji and {Kokorev}, Vasily and {Dayal}, Pratika and {Nanayakkara}, Themiya and {Williams}, Christina C. and {Marchesini}, Danilo and {Zitrin}, Adi and {van Dokkum}, Pieter},
        title = "{UNCOVER: JWST Spectroscopy of Three Cold Brown Dwarfs at Kiloparsec-scale Distances}",
      journal = {\apj},
         year = 2024,
        month = feb,
       volume = {962},
       number = {2},
          eid = {177},
        pages = {177}
}

@ARTICLE{2025A&A...697A.189D,
       author = {{de Graaff}, Anna and {Brammer}, Gabriel and {Weibel}, Andrea and {Lewis}, Zach and {Maseda}, Michael V. and {Oesch}, Pascal A. and {Bezanson}, Rachel and {Boogaard}, Leindert A. and {Cleri}, Nikko J. and {Cooper}, Olivia R. and {Gottumukkala}, Rashmi and {Greene}, Jenny E. and {Hirschmann}, Michaela and {Hviding}, Raphael E. and {Katz}, Harley and {Labb{\'e}}, Ivo and {Leja}, Joel and {Matthee}, Jorryt and {McConachie}, Ian and {Miller}, Tim B. and {Naidu}, Rohan P. and {Price}, Sedona H. and {Rix}, Hans-Walter and {Setton}, David J. and {Suess}, Katherine A. and {Wang}, Bingjie and {Whitaker}, Katherine E. and {Williams}, Christina C.},
        title = "{RUBIES: A complete census of the bright and red distant Universe with JWST/NIRSpec}",
      journal = {\aap},
         year = 2025,
        month = may,
       volume = {697},
          eid = {A189},
        pages = {A189},
 primaryClass = {astro-ph.GA}
}

@ARTICLE{2024Sci...384..890H,
       author = {{Heintz}, Kasper E. and {Watson}, Darach and {Brammer}, Gabriel and {Vejlgaard}, Simone and {Hutter}, Anne and {Strait}, Victoria B. and {Matthee}, Jorryt and {Oesch}, Pascal A. and {Jakobsson}, P{\'a}ll and {Tanvir}, Nial R. and {Laursen}, Peter and {Naidu}, Rohan P. and {Mason}, Charlotte A. and {Killi}, Meghana and {Jung}, Intae and {Hsiao}, Tiger Yu-Yang and {Abdurro'uf} and {Coe}, Dan and {Arrabal Haro}, Pablo and {Finkelstein}, Steven L. and {Toft}, Sune},
        title = "{Strong damped Lyman-{\ensuremath{\alpha}} absorption in young star-forming galaxies at redshifts 9 to 11}",
      journal = {Science},
         year = 2024,
        month = may,
       volume = {384},
       number = {6698},
        pages = {890-894}
}

@ARTICLE{2010ApJS..190..100K,
       author = {{Kirkpatrick}, J. Davy and {Looper}, Dagny L. and {Burgasser}, Adam J. and {Schurr}, Steven D. and {Cutri}, Roc M. and {Cushing}, Michael C. and {Cruz}, Kelle L. and {Sweet}, Anne C. and {Knapp}, Gillian R. and {Barman}, Travis S. and {Bochanski}, John J. and {Roellig}, Thomas L. and {McLean}, Ian S. and {McGovern}, Mark R. and {Rice}, Emily L.},
        title = "{Discoveries from a Near-infrared Proper Motion Survey Using Multi-epoch Two Micron All-Sky Survey Data}",
      journal = {\apjs},
         year = 2010,
        month = sep,
       volume = {190},
       number = {1},
        pages = {100-146},
 primaryClass = {astro-ph.SR}
}

@ARTICLE{2024A&A...689A..73M,
       author = {{Maseda}, Michael V. and {de Graaff}, Anna and {Franx}, Marijn and {Rix}, Hans-Walter and {Carniani}, Stefano and {Laseter}, Isaac and {Dudzevi{\v{c}}i{\={u}}t{\.{e}}}, Ugn{\.{e}} and {Rawle}, Tim and {Parlanti}, Eleonora and {Arribas}, Santiago and {Bunker}, Andrew J. and {Cameron}, Alex J. and {Charlot}, Stephane and {Curti}, Mirko and {D'Eugenio}, Francesco and {Jones}, Gareth C. and {Kumari}, Nimisha and {Maiolino}, Roberto and {{\"U}bler}, Hannah and {Saxena}, Aayush and {Smit}, Renske and {Willott}, Chris and {Witstok}, Joris},
        title = "{The NIRSpec Wide GTO Survey}",
      journal = {\aap},
         year = 2024,
        month = sep,
       volume = {689},
          eid = {A73},
        pages = {A73}
}

@ARTICLE{2023ApJ...946L..13F,
       author = {{Finkelstein}, Steven L. and {Bagley}, Micaela B. and {Ferguson}, Henry C. and {Wilkins}, Stephen M. and {Kartaltepe}, Jeyhan S. and {Papovich}, Casey and {Yung}, L.~Y. Aaron and {Arrabal Haro}, Pablo and {Behroozi}, Peter and {Dickinson}, Mark and {Kocevski}, Dale D. and {Koekemoer}, Anton M. and {Larson}, Rebecca L. and {Le Bail}, Aur{\'e}lien and {Morales}, Alexa M. and {P{\'e}rez-Gonz{\'a}lez}, Pablo G. and {Burgarella}, Denis and {Dav{\'e}}, Romeel and {Hirschmann}, Michaela and {Somerville}, Rachel S. and {Wuyts}, Stijn and {Bromm}, Volker and {Casey}, Caitlin M. and {Fontana}, Adriano and {Fujimoto}, Seiji and {Gardner}, Jonathan P. and {Giavalisco}, Mauro and {Grazian}, Andrea and {Grogin}, Norman A. and {Hathi}, Nimish P. and {Hutchison}, Taylor A. and {Jha}, Saurabh W. and {Jogee}, Shardha and {Kewley}, Lisa J. and {Kirkpatrick}, Allison and {Long}, Arianna S. and {Lotz}, Jennifer M. and {Pentericci}, Laura and {Pierel}, Justin D.~R. and {Pirzkal}, Nor and {Ravindranath}, Swara and {Ryan}, Russell E. and {Trump}, Jonathan R. and {Yang}, Guang and {Bhatawdekar}, Rachana and {Bisigello}, Laura and {Buat}, V{\'e}ronique and {Calabr{\`o}}, Antonello and {Castellano}, Marco and {Cleri}, Nikko J. and {Cooper}, M.~C. and {Croton}, Darren and {Daddi}, Emanuele and {Dekel}, Avishai and {Elbaz}, David and {Franco}, Maximilien and {Gawiser}, Eric and {Holwerda}, Benne W. and {Huertas-Company}, Marc and {Jaskot}, Anne E. and {Leung}, Gene C.~K. and {Lucas}, Ray A. and {Mobasher}, Bahram and {Pandya}, Viraj and {Tacchella}, Sandro and {Weiner}, Benjamin J. and {Zavala}, Jorge A.},
        title = "{CEERS Key Paper. I. An Early Look into the First 500 Myr of Galaxy Formation with JWST}",
      journal = {\apjl},
         year = 2023,
        month = mar,
       volume = {946},
       number = {1},
          eid = {L13},
        pages = {L13}
}

@ARTICLE{2025ApJS..277....4D,
       author = {{D'Eugenio}, Francesco and {Cameron}, Alex J. and {Scholtz}, Jan and {Carniani}, Stefano and {Willott}, Chris J. and {Curtis-Lake}, Emma and {Bunker}, Andrew J. and {Parlanti}, Eleonora and {Maiolino}, Roberto and {Willmer}, Christopher N.~A. and {Jakobsen}, Peter and {Robertson}, Brant E. and {Johnson}, Benjamin D. and {Tacchella}, Sandro and {Cargile}, Phillip A. and {Rawle}, Tim and {Arribas}, Santiago and {Chevallard}, Jacopo and {Curti}, Mirko and {Egami}, Eiichi and {Eisenstein}, Daniel J. and {Kumari}, Nimisha and {Looser}, Tobias J. and {Rieke}, Marcia J. and {Rodr{\'\i}guez Del Pino}, Bruno and {Saxena}, Aayush and {{\"U}bler}, Hannah and {Venturi}, Giacomo and {Witstok}, Joris and {Baker}, William M. and {Bhatawdekar}, Rachana and {Bonaventura}, Nina and {Boyett}, Kristan and {Charlot}, Stephane and {Danhaive}, A. Lola and {Hainline}, Kevin N. and {Hausen}, Ryan and {Helton}, Jakob M. and {Ji}, Xihan and {Ji}, Zhiyuan and {Jones}, Gareth C. and {Juod{\v{z}}balis}, Ignas and {Maseda}, Michael V. and {P{\'e}rez-Gonz{\'a}lez}, Pablo G. and {Perna}, Michele and {Pusk{\'a}s}, D{\'a}vid and {Shivaei}, Irene and {Silcock}, Maddie S. and {Simmonds}, Charlotte and {Smit}, Renske and {Sun}, Fengwu and {Villanueva}, Natalia C. and {Williams}, Christina C. and {Zhu}, Yongda},
        title = "{JADES Data Release 3: NIRSpec/Microshutter Assembly Spectroscopy for 4000 Galaxies in the GOODS Fields}",
      journal = {\apjs},
         year = 2025,
        month = mar,
       volume = {277},
       number = {1},
          eid = {4},
        pages = {4}
}

@ARTICLE{2025ApJ...983...18R,
       author = {{Roberts-Borsani}, Guido and {Bagley}, Micaela and {Rojas-Ruiz}, Sof{\'\i}a and {Treu}, Tommaso and {Morishita}, Takahiro and {Finkelstein}, Steven L. and {Trenti}, Michele and {Arrabal Haro}, Pablo and {Ba{\~n}ados}, Eduardo and {Ch{\'a}vez Ortiz}, {\'O}scar A. and {Chworowsky}, Katherine and {Hutchison}, Taylor A. and {Larson}, Rebecca L. and {Leethochawalit}, Nicha and {Leung}, Gene C.~K. and {Mason}, Charlotte and {Somerville}, Rachel S. and {Stiavelli}, Massimo and {Yung}, L.~Y. Aaron and {Kassin}, Susan A. and {Soto}, Christian},
        title = "{The BoRG-JWST Survey: Program Overview and First Confirmations of Luminous Reionization-era Galaxies from Pure-parallel Observations}",
      journal = {\apj},
         year = 2025,
        month = apr,
       volume = {983},
       number = {1},
          eid = {18},
        pages = {18}
}

@ARTICLE{2024ApJ...974...92B,
       author = {{Bezanson}, Rachel and {Labbe}, Ivo and {Whitaker}, Katherine E. and {Leja}, Joel and {Price}, Sedona H. and {Franx}, Marijn and {Brammer}, Gabriel and {Marchesini}, Danilo and {Zitrin}, Adi and {Wang}, Bingjie and {Weaver}, John R. and {Furtak}, Lukas J. and {Atek}, Hakim and {Coe}, Dan and {Cutler}, Sam E. and {Dayal}, Pratika and {van Dokkum}, Pieter and {Feldmann}, Robert and {F{\"o}rster Schreiber}, Natascha M. and {Fujimoto}, Seiji and {Geha}, Marla and {Glazebrook}, Karl and {de Graaff}, Anna and {Greene}, Jenny E. and {Juneau}, St{\'e}phanie and {Kassin}, Susan and {Kriek}, Mariska and {Khullar}, Gourav and {Maseda}, Michael and {Mowla}, Lamiya A. and {Muzzin}, Adam and {Nanayakkara}, Themiya and {Nelson}, Erica J. and {Oesch}, Pascal A. and {Pacifici}, Camilla and {Pan}, Richard and {Papovich}, Casey and {Setton}, David J. and {Shapley}, Alice E. and {Smit}, Renske and {Stefanon}, Mauro and {Taylor}, Edward N. and {Williams}, Christina C.},
        title = "{The JWST UNCOVER Treasury Survey: Ultradeep NIRSpec and NIRCam Observations before the Epoch of Reionization}",
      journal = {\apj},
         year = 2024,
        month = oct,
       volume = {974},
       number = {1},
          eid = {92},
        pages = {92}
}

@ARTICLE{2023ApJ...948L..18N,
       author = {{Nelson}, Erica J. and {Suess}, Katherine A. and {Bezanson}, Rachel and {Price}, Sedona H. and {van Dokkum}, Pieter and {Leja}, Joel and {Wang}, Bingjie and {Whitaker}, Katherine E. and {Labb{\'e}}, Ivo and {Barrufet}, Laia and {Brammer}, Gabriel and {Eisenstein}, Daniel J. and {Gibson}, Justus and {Hartley}, Abigail I. and {Johnson}, Benjamin D. and {Heintz}, Kasper E. and {Mathews}, Elijah and {Miller}, Tim B. and {Oesch}, Pascal A. and {Sandles}, Lester and {Setton}, David J. and {Speagle}, Joshua S. and {Tacchella}, Sandro and {Tadaki}, Ken-ichi and {{\"U}bler}, Hannah and {Weaver}, John. R.},
        title = "{JWST Reveals a Population of Ultrared, Flattened Galaxies at 2 {\ensuremath{\lesssim}} z {\ensuremath{\lesssim}} 6 Previously Missed by HST}",
      journal = {\apjl},
         year = 2023,
        month = may,
       volume = {948},
       number = {2},
          eid = {L18},
        pages = {L18}
}
\bibliographystyle{aasjournal}

\end{document}